\documentstyle{general}
\def\pom{{I\!\!P}}            
\begin{document}
\title{Low $x$ physics and Structure Functions }
\author{Rahul Basu
\thanks{Presented at the Fifth Workshop on  High Energy
Physics Phenomenology, 12 Jan 1998 -- 25 Jan 1998 at Pune, India}
\affiliation{The Institute of Mathematical Sciences, Chennai (Madras) 
600 113, India}
}

\abstract{Abstract}{
In this talk I review the behaviour of structure functions at low
values of Bjorken $x$ and discuss the theoretical underpinnings with
particular attention to resummation schemes. I present the need for
'less inclusive' events to distinguish between various resummation
schemes and discuss the various difficulties in differentiating
experimentally between different schemes. } 

\maketitle
\section{Introduction}
The wealth of data on structure functions at low $x$ that has emerged
from HERA in the last six years or so have provided fresh and
interesting challenges to theorists to explain the steep rise in the
structure function $F_2(x) $ along with many other new aspects of
perturbative QCD at these hitherto unprobed kinematic regions. In this
talk I will concentrate on structure functions and discuss in some
detail the experimental data along with the theoretical attempts made to
explain them. I will not be touching upon many of the other interesting
aspects of low $x$ QCD {\em viz} Large Rapidity Gap Events, Quarkonia
Production, spin physics, Photon Structure Functions and so on, most
of which will be addressed by other speakers in this workshop. 

The basic interaction picture is the usual DIS diagram for
the process
$$
e(k) + p(p) \rightarrow e(k^\prime) + X.
$$

The following kinematic variables completely describe the process:
\begin{eqnarray}
s=(k+p)^2 & \simeq & 4E_eE_p \\
Q^2 = -q^2 & \simeq & 2E_eE_{e^\prime}(1+cos\theta) \\
y=\frac{p.q}{p.k} & \simeq & 1 - \frac{E_{e^\prime}}{2E_e}(1-cos\theta) \\
x=\frac{Q^2}{2p.q}& \simeq & \frac{Q^2}{ys} \\
W=(q+p)^2 & \simeq & -Q^2 + ys 
\end{eqnarray}
here $s$ is the CM energy squared, $E_e(E_e^\prime)$ is the energy of the
initial (scattered) electron, $\theta$ the angle of the scattered electron in
the lab frame with respect to the proton direction, $Q^2$ is the negative of
the momentum transfer squared to the proton, $y$ is the fraction of the
electron energy carried by the virtual photon in the proton rest frame, the
Bjorken variable $x$ is the fraction of the proton energy carried by the
struck quark and $W$ is the hadronic invariant mass. The masses of all
particles are neglected.

At HERA, a 27 GeV electron beam collides head-on with an 820 GeV
proton beam giving 
$$
s \simeq 4E_eE_p \sim 10^5 GeV^2
$$
which is much larger than hitherto obtained at fixed target
experiments. (Since 1994, the electron beam has been replaced by a
positron beam at the same energy). As a result the two experiments H1 
and ZEUS can measure the
structure function $F_2$ in a completely new $(x,Q^2)$ domain (see
Figure 1).
\begin{figure}
\figurebox{}{8truecm}[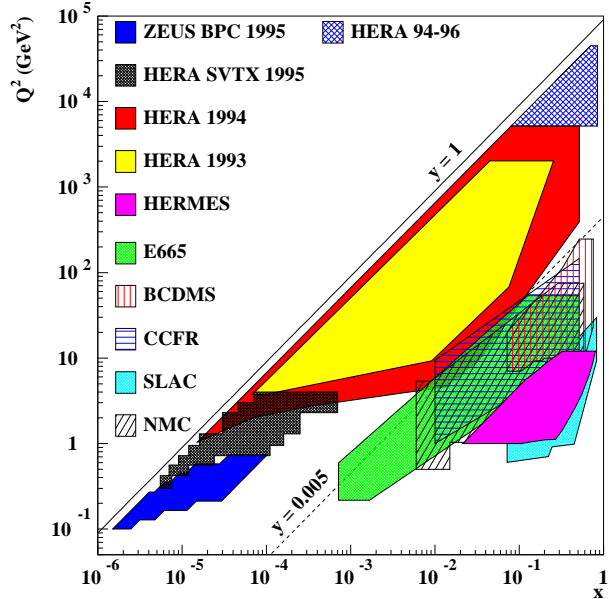]
\caption{Phase space coverage of the $F_2$ measurements} 
\end{figure}

The Born cross section for single photon exchange in DIS is given by
\begin{equation} 
\frac{d^2\sigma}{dxdQ^2}=\frac{2\pi\alpha^2}{Q^2x}\left[2(1-y)+\frac{y^2}
{1+R}\right]F_2(x,Q^2)
\end{equation} 
where $R$ is the photoproduction cross section ratio for longitudinal
and transverse polarised photon, $R=\sigma_L/\sigma_T$. $R$ has not
been measured at HERA; therefore a QCD prescription based on
parametrisation of parton densities is used. $Q^2$ is directly
measured from the scattered electron but $x$ is calculated from $Q^2$
and $y$ and therefore depends on the experimental resolution of $y$.

\section{Measurement of $F_2(x,Q^2)$}
The data for $F_2$ from H1 and ZEUS \cite{h1,zeus} is shown in 
figures 2, 3, and 4, along with some of the older fixed target data. It is clear
that at fixed $Q^2$, $F_2$ rises with decreasing $x$ down to the
smallest $Q^2$, the steepness of the rise decreasing with decreasing
$Q^2$. Similarly at fixed $x < 0.1$ $F_2$ rises with $Q^2$, the rise
becoming steeper as $x$ decreases. In fact, as seen in figure 2, upto
$Q^2 \sim 0.85 GeV^2$ the data favour a parametrisation based on a
soft pomeron  such as that suggested by Donnachie and Landshoff
\cite{dola}. For larger values of $Q^2$ ($Q^2 > 1.0 GeV^2$) the usual
QCD based parametrisations of GRV or MRS \cite{grv, mrs} give a good
description of the data, as seen in Figure 3. 
\begin{figure}[htb]
\figurebox{}{8truecm}[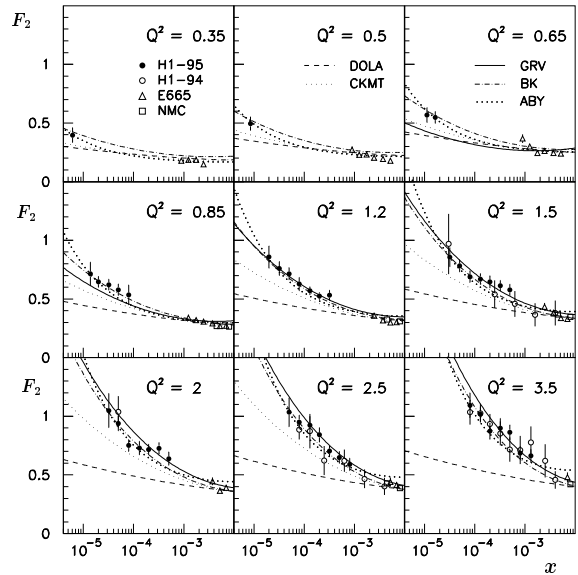]
\caption{$F_2(x,Q^2)$ as a function of $x$ for low $Q^2$. 
The Donnachie Landshoff (DOLA) and the standard QCD evolution
represented by GRV are shown}
\label{Fig 2.}
\end{figure}
\begin{figure}[htb]   
\figurebox{}{8truecm}[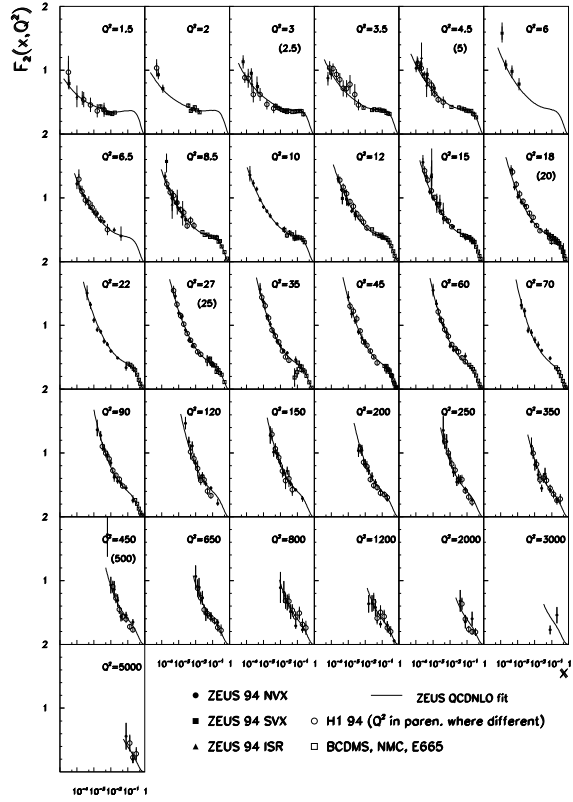]
\caption{$F_2(x,Q^2)$ as a function of $x$ for higher $Q^2$
compared with model predictions.}
\label{Fig 3.}   
\end{figure} 
\begin{figure}[htb]   
\figurebox{}{8truecm}[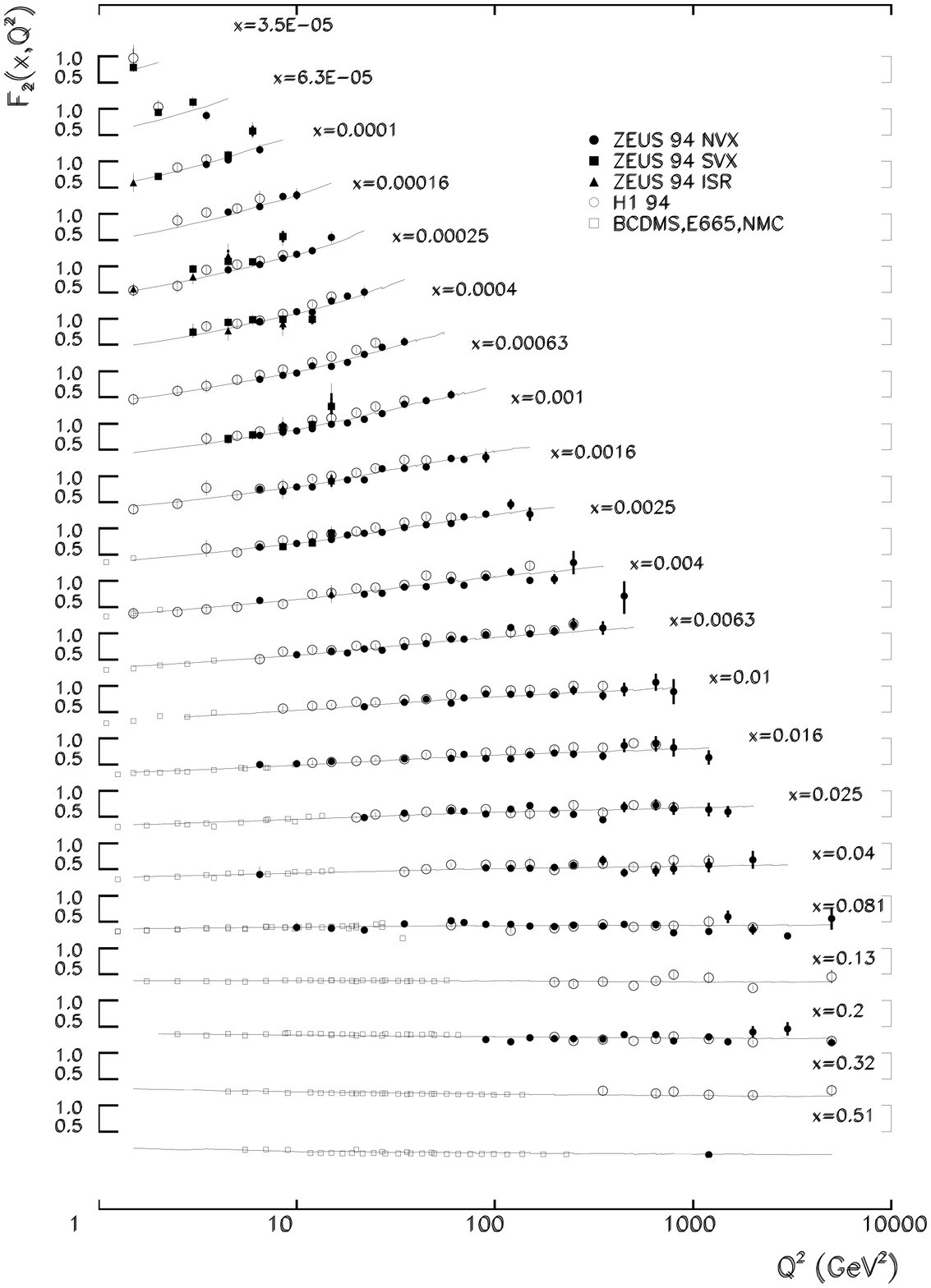]
\caption{Compilation of $F_2$ measurements as a function $Q^2$ for
selected values of $x$. The curves are a NLO QCD fit by H1. }
\end{figure} 

The behaviour of $F_2$ as a function of $Q^2$ for selected values of
$x$ is presented in figure 4. The growth of the structure function
with $Q^2$ for $x<0.01$ over three orders of magnitude in $Q^2$ is
clearly seen. With the new data from HERA the gap between fixed
target and the HERA experiments has been filled and the remarkable
agreement
with a NLO QCD fit performed by H1 over three decades in $Q^2$ and
over four decades in $x$ is clearly  seen. 

The H1 experiment has parametrised the rise of $F_2$ with $x$ by $F \sim
x^{-\lambda}$ at fixed $Q^2$ values. The result for $\lambda$ as a
function of $Q^2$ is shown in Figure 5. 
It is clear from the figure that $\lambda$ increases with increasing
$Q^2$ reaching a value of around $0.4-0.5$ around $Q^2 \sim 10^2-10^3
GeV^2$.
\begin{figure}[htb]   
\figurebox{}{8truecm}[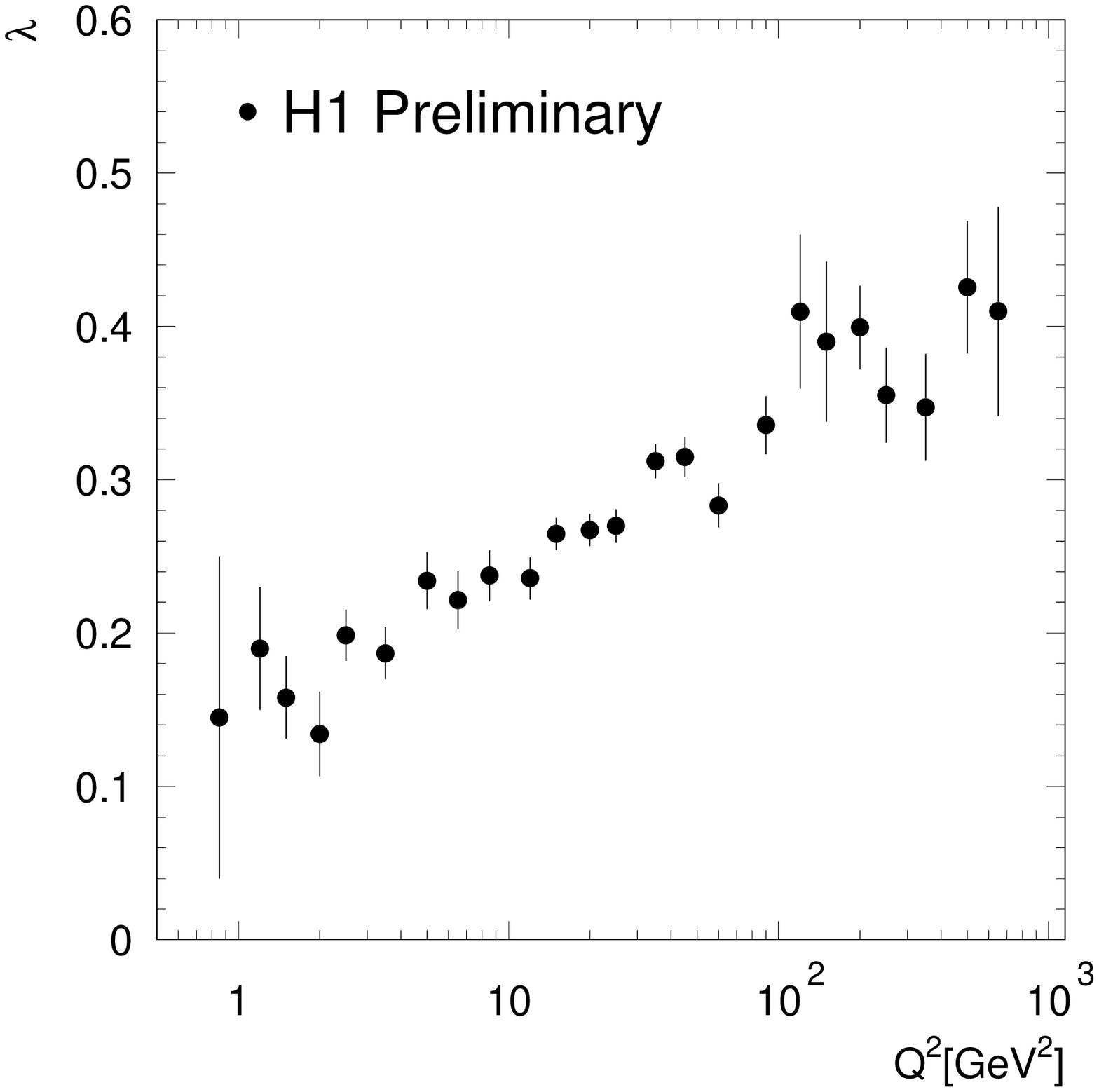]
\caption{Variation of the exponent $\lambda$ from fits of the form
$F_2 \sim x^{-\lambda}$ at fixed $Q^2$ values and $x < 0.1$.
}
\end{figure} 

\section{The behaviour of $F_2(x)$}
The general behaviour of $F_2(x)$ can be understood as follows. Below
$Q^2\simeq 0.85 GeV^2 $ the Regge theory predictions for a soft pomeron
like behaviour seems to work quite well \cite{dola}. In this approach,
the total cross section (for ${\bar p}p$, $K^\pm p$, $\pi^\pm p$ etc.
are fitted to a form
\begin{equation} 
\sigma_{tot}= As^\epsilon + B s^{-\eta}
\end{equation}
where $\epsilon=0.08$ corresponds to a soft pomeron intercept
$\alpha_{\pom}(0)=1+\epsilon$  and $\eta=0.45$ corresponds to the $\rho$
meson trajectory intercept $\alpha=1+\eta$. Assuming a dipole model for
the proton, the $x$ dependence of $F_2$ can be extracted;
\begin{equation} 
F_2^{DOLA}(x,Q^2)=Ax^{-0.08}\frac{Q^2}{Q^2+a^2}+Bx^{0.45}\frac{Q^2}{Q^2+b^2}
\end{equation}
Thus the DOLA analysis predicts a gentle rise of $x^{-0.08}$ for $F_2$ with 
decreasing $x$. 

However this prediction, as we have seen from the data fails for $Q^2 >
0.85 GeV^2$ where standard QCD evolution takes over, given
generically by the GRV fits. For the rest of the section we shall be
discussing the nature of this 'good' agreement of the data with QCD
evolution. 

In QCD, the structure function $F_2(x,Q^2)$ is given in terms of the parton 
distributions
\begin{equation}
\frac{F_2(x,Q^2)}{x}\equiv\sum_{i=1}^{n_f}e_i^2C_i\otimes (q_i+\bar q_i)+C_g
\otimes g.
\end{equation}
The evolution of the structure function $F_2$ as a function of $Q^2$ is 
governed, in
perturbative QCD, by the Dokshitser-Gribov-Lipatov-Altarelli-Parisi
(DGLAP) equation \cite{dok, glap}, 
\begin{equation}
{\partial \over \partial \ln Q^2}
 \left(\begin{array}{c} q \\ g \end{array}\right)
 = {\alpha_s(Q^2) \over 2 \pi} \left[\begin{array}{cc}
 P_{qq} & P_{qg}  \\
 P_{gq} & P_{gg}
\end{array}\right] \otimes
\left(\begin{array}{c}q \\ g\end{array}\right) \, ,
\end{equation}
Here $q$ and $g$ are the quark and gluon distributions, $\otimes$ denotes 
convolution
with respect to $x$, i.e.$[f\otimes g](x)\equiv\int_x^1 {dy \over y}f({x\over
y})g(y)$, $n_f$ is the number of flavours of quarks, $e_i$ is the electric 
charge of
the quark $q_i$. The coefficient functions $C$'s at leading order are
\begin{equation}
C_i(x,Q^2)=\delta(1-x);\ \ \ C_g(x,Q^2)=0.
\end{equation}
At higher orders, they depend on the specific factorization scheme. In
the common parton scheme the above remains true to all orders. 

The splitting functions have the expansion
\begin{equation}
P_{ij}(x,Q^2)= \frac{\alpha_s}{2\pi} P_{ij}^{(1)}(x)+
\left(\frac{\alpha_s}{2\pi}\right)^2 P_{ij}^{(2)}(x)+ \ldots \, .
\end{equation}
The first two terms above define the NLO DGLAP evolution. 

The rise of $F_2$ as a function of $x$ is explained by the usual DGLAP 
evolution and
also by more non-conventional dynamics like the BFKL evolution \cite{bfkl}. 
Let us briefly review the two different approaches.

Perturbative QCD does not provide any information on hadronic structure.
However the idea of factorisation comes to the rescue. Very briefly, the
electron proton cross section can be written as a convolution of the
partonic cross section of the partons in the proton scattering off the
electron times the probability of finding the parton in the proton. 
Symbolically
\begin{equation}
\sigma_{ep}=\sum_i \big[ f_{i/p}\otimes \sigma_{ei}\big]
\end{equation}
where $\sigma_{ei}$ the partonic cross section is calculable in
perturbative QCD. Unfortunately this cross section is plagued with 
infra red divergences which are of two kinds - soft divergences
($k\rightarrow 0$) which are cancelled by the virtual diagrams and
collinear divergences ($k_T\rightarrow 0$) which are absorbed
(factorised) into the bare parton distributions. The result of this last
step is to get new redefined parton distributions which depend now on
the factorisation scale $\mu_F$. Thus the IR safe cross section for $ep$
has the form
\begin{equation}
\sigma_{ep}=\sum_i \big[ f_{i/p}(\mu_F^2)\otimes \sigma_{ei}(\mu_F^2)\big]
\end{equation}
The scale $\mu_F$ is arbitrary but usually taken to be equal to $Q^2$ in
which case the lowest order contribution to $\sigma_{ei}$ is the Born
graph. Then $f_{i/p}(Q^2)$ is the parton density in the proton as seen by
a photon of virtuality $Q^2$. 

Since $\sigma_{ei}(\mu_F^2)$ can be calculated perturbative for any
scale, the change in the redefined parton distribution can also be
calculated. Thus while perturbative QCD does not predict the form of
$f_{i/p}$, it does predict the change of $f_{i/p}$ with scale. These are
given by the various evolution equations, the differences between them
being the approximations that are used to restrict the phase space for
radiation. These approximations are then valid in regions of $x,Q^2$
where the selected contributions are the dominant ones. These therefore
give different evolution equations (D)GLAP, DLL, BFKL \ldots. 

To understand the leading log summation technique in the DGLAP approach, 
we use the Dokshitzer method \cite{dok}. At low values of $x$, the gluon
is the dominant parton in the proton. He showed that the GLAP summation 
of the $(\alpha_s \log
Q^2)^n $ terms amounts to a sum of gluon ladder diagrams of the type shown 
in Figure
6 (a) with $n$ gluon rungs. The form of the gluon splitting function 
$P_{gg}\sim 6/x$
for small $x$ gives the evolution of the gluon distribution function as
\begin{eqnarray}
xg(x,Q^2)&=&\sum_n(\frac{3\alpha_s}{\pi})^n\int^{Q^2}\frac{dk_{nT}^2}{k_{nT}^2}
\ldots
\int^{k_{3T}^2}\frac{dk_{2T}^2}{k_{2T}^2}\int^{k_{2T}^2}\frac{dk_{1T}^2}
{k_{1T}^2} \nonumber \\
&&\times \int_x^1\frac{d\xi_n}{\xi_n}\ldots\int_{\xi_3}^1\frac{\xi_2}{\xi_2}
\int_{\xi_2}^1\frac{\xi_1}{\xi_1}\xi_1g(\xi_1,Q_0^2)
\end{eqnarray}
\begin{figure}[htb]   
\figurebox{}{8truecm}[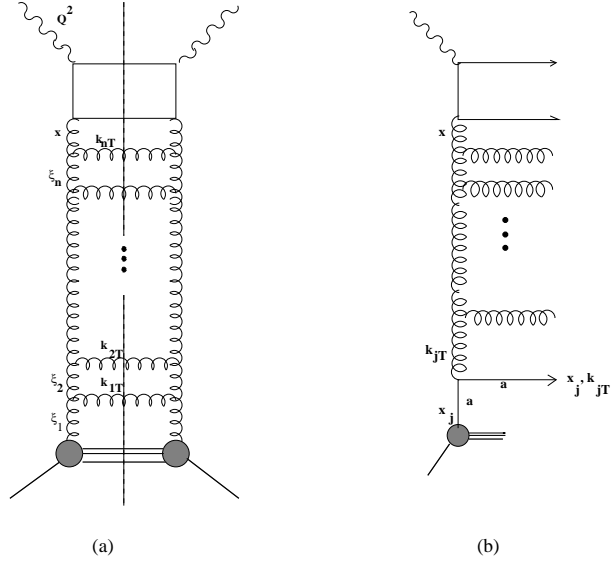]
\caption{A gluon ladder diagram that contributes to the DLL or BFKL
ladder summations. }
\end{figure} 
It is clear that the $(\log Q^2)^n$ builds up from the nested $k$ integrations. 
In fact this contribution comes from a region where the transverse momentum of
the emitted gluons are strongly ordered
\begin{equation}
Q^2 >> k_{nT}^2 >> \ldots >> k_{2T}^2 >> k_{1T}^2
\end{equation}
This kinematic region is therefore relevant when $Q^2$ is large but $x$
is not so small that $\alpha_s(Q^2)\ln (1/x)$ is large i.e
$$
\alpha_s(Q^2)\ln (1/x)<< \alpha_s(Q^2)\ln(Q^2/Q_0^2) < 1
$$
Similarly, the $(\log(1/x))^n$ term comes from the region where the longitudinal
momentum fractions are strongly ordered
\begin{equation}
x << \xi_n <<\ldots << \xi_2 << \xi_1.
\end{equation}
In the region of small $x$ the $\log(1/x)$ terms have also to be summed. 
With the assumption of $\xi_1g(\xi_1,Q_0^2)$ approaching a constant, 
say $G_0$ the result we get is
\begin{eqnarray}
xg(x,Q^2)& \sim &\sum_n\left(\frac{3\alpha_s}{\pi}\right)^n\left\{{1\over n!}
\left[\ln\left(
{Q^2\over Q_0^2}\right)\right]^n\right\}\left\{{1\over n!}\left[\ln\left(
{1\over x}\right)\right]^nG_0\right\} \\ \nonumber 
&& \sim G_0\exp\left\{2\left[\frac{3\alpha_s}{\pi}\ln\left(Q^2\over Q_0^2\right)
\ln \left(1\over x\right)\right]^{1\over 2}\right\}
\end{eqnarray}
in the limit of large $Q^2$ and small $x$.

This is an example of an all order double leading log (DLL) summation of 
$\alpha_s\log Q^2
\log(1/x)$ obtained by summing the strongly ordered gluon ladder diagrams 
as we have
just explained. It tells us that as $x$ decreases, $xg(x,Q^2)$ increases faster than 
any power of $\log(1/x)$ but slower than any power of $1/x$. It is valid
in a region where 
$$
\left\{ \begin{array}{l}
      \alpha_s(Q^2)\ln(Q^2/Q_0^2)\\
      \alpha_s(Q^2)\ln(1/x)
      \end{array}
      \right\} < < \alpha_s(Q^2)\ln(Q^2/Q_0^2)\ln(1/x) < 1
$$

If $k_{iT}\approx k_{i+1T}$ then we lose one power of $\log Q^2$. However, 
at small
$x$, this can be compensated by a large $\log(1/x)$ factor. Relaxing the strong
ordering constraint on the $k_T$'s and summing just the $\log(1/x)$ in the 
small $x$
region will give us the leading log summation in $\log(1/x)$ instead of DLL 
and we get
\begin{eqnarray}
xg(x,Q^2)& \approx& \sum_n(\frac{3\alpha_s}{\pi})^n\frac{1}{n!}\left[
c\ \log\left( {1\over x}\right)\right]^n \nonumber \\
&& \sim \exp\left[\lambda \log\left({1\over x}\right)\right] \nonumber \\
&&  \sim x^{-\lambda}
\end{eqnarray}
where $\lambda=(3\alpha_s/\pi)c$. This difficult summation was done 
rigorously by
Balitskij, Fadin, Kuraev and Lipatov (BFKL) \cite{bfkl}. The constant 
$c=4\ \log2$ and
so, for $\alpha_s \approx 0.2$, $\lambda\approx 0.5$. We have highly 
simplified the
discussion of the BFKL equation, and in practise one needs to work 
in terms of the
unintegrated gluon distribution $f(x,k_T^2)$ and integrate $f(x,k_T^2)/k_T^2$ 
to get the final gluon distribution. The BFKL region is therefore one
where $x$ is small and $Q^2$ is not large enough to reach the DLL
regime. 

We thus have two different predictions for the rise of $F_2$, the 
DLL (from the DGLAP
evolution equations) and the BFKL. In principle it should therefore be 
possible to
distinguish between these predictions by looking at the $F_2$ data from HERA. 
Unfortunately this is not that simple.  The steepness of the rise of $F_2$ with
decreasing $x$ can be controlled by varying $Q_0^2$, or the starting 
distribution 
$g(\xi_1,Q_0^2)$. In addition, the set of equations based only on 
$F_2$ measurements
is underconstrained and one needs one other measurement like the longitudinal
structure function $F_L$ in order to constrain the system fully. 

The net consequence of the above is that the present data on $F_2$ does not
distinguish betweem the BFKL and DGLAP (or for that matter the CCFM which 
embodies both) predictions. 

One of the reasons that the data on $F_2$ is not sufficient is due to a
phenomenon called $k_T$ diffusion. Let us explain this in some detail.
The BFKL equation predicts the
behaviour of the unintegrated gluon distribution $f(x,k_T^2)$ to be
\begin{equation}
\frac{f(x,k_T^2)}{\sqrt{k_T^2}}\propto \frac{(x/x_0)^{-\lambda}}
{\sqrt{2\pi\lambda^"\ln(x_0/x)}}\exp\left[-\frac{\ln^2(k_T^2/\bar{k_T^2})}
{2\lambda^"\ln(x_0/x)}\right]
\end{equation}
for fixed $\alpha_s$. Here $\lambda={3\alpha_s\over \pi}4\ln 2\simeq
0.5$ for $\alpha_s = 0.19$, $\lambda^" = {3\alpha_s\over
\pi}28\zeta(3)$. Thus the unintegrated gluon distribution exhibits a
Gaussian in $\ln k_T^2$ with a width which increases with the BFKL
evolution length $\sqrt{\ln(x_0/x)}$. This form implies that individual
evolution paths follow a kind of randon walk in $k_T^2$ and an ensemble
of paths exhibits a diffusion pattern according to a Gaussian in $\ln
k_T^2$. The net consequence of this is that $k_T^2$ can diffuse into a
region of very small values where perturbation theory no longer holds.
Contrast this with $k_T$ ordering in DGLAP which ensures that the
$k_T$'s along a ladder are always contrained to remain within a 
region sufficiently removed from the infrared. 
The calculation of $F_2$ involving an 
integration over all $k_T^2$ in the BFKL formalism could therefore include 
diffusion of $k_T$
into the IR region  where perturbation theory fails. A possible solution
is to study special final state configurations where diffusion into the
IR region is minimised by fixing the start and end point of evolution
for the above region. 

This is the idea behind a suggestion by Mueller
\cite{mueller} on measuring an observable that is less inclusive than the $F_2$
measurement (in which none of the properties of the hadronic final state is
measured). He suggests studying DIS events at small $x$ containing an identified
high-energy jet emitted close to the jet of proton fragments (see Figure 6(b)). The
identified jet originates from the parton labelled $a$. If we study events with $x_j$
large ($\ge 0.05$) and $x$ very small ($\approx 10^{-4}$) the ratio $x/x_j$ 
will be sufficiently small to reveal the $(x/x_j)^{-\lambda}$ behaviour of the 
BKFL ladder. Details may be found in the references mentioned above. This study
is currently under way and preliminary results are reported in 
\cite{H1-forward}.
We will have more to say on this later.

The behaviour of $F_2$ as a function of both $Q^2$ and $x$ (for large $Q^2$
and small $x$) can be neatly combined through the double
asymptotic scaling analysis of Ball and Forte \cite{bf}. They build on
an old analysis of de Rujula et al. \cite{derujula} to show that the 
gluon distribution function for decreasing $x$ increases faster than a 
power of $\ln (1/x)$ but slower than a power of $1/x$.  
There are numerous other evolution equations and details of the current
phenomenology of these evolution equations can be found in
\cite{coopersarkar}

One of the by-products of the measurement of the structure function
$F_2$ and its $\log Q^2$ evolution is that the gluon distribution inside
the proton has been measured with greater accuracy ({\it cf.} Figure 14
in \cite{foster}). 

To summarise this section, the structure function $F_2$ has been found
to rise steeply with decreasing $x$ down to $Q^2$ as low as $0.85
GeV^2$. This rise can be parametrised by a form $x^{-\lambda}$ where
$\lambda$ is found to be a function of $Q^2$ and reaches a value of
around $0.4-0.5$ around $1000 GeV^2$. This behaviour is described very
well by conventional NLO QCD (through the DGLAP equation). However this
does not rule out the existence of non-conventional dynamics like BFKL
because the system is underconstrained. Thus some other measurement
which is less inclusive than $F_2$ like the forward jet mentioned above,
or a measurement of the longitudinal structure function $F_L$ is
required. In any case, some new physics must take over at sufficiently
low $x$ otherwise the indefinite rise of $F_2$ would violate the
unitarity bound.

\section{Longitudinal Structure Function}

The longitudinal structure function $F_L = F_2 - 2xF_1$ is related to
the differential cross section by
\begin{equation}
\frac{d^2\sigma}{dxdQ^2}=\frac{4\pi\alpha^2}{xQ^4}\left[(1-y+\frac{y^2}{2})
F_2 - \frac{y^2}{2} F_L\right]
\end{equation}
$F_L$ is related to $F_2$ and the gluon distribution function through
the Altarelli-Martinelli equation
\begin{equation}
F_L(x,Q^2)=\frac{\alpha_s}{4\pi}x^2\int_x^1\frac{dz}{z^3}
\left[\frac{16}{3}F_2(z,Q^2)+8\sum_ie_i^2(1-x/z)zg(z,Q^2)\right]
\end{equation}
At low $x$ the dominant term is the second term. 

A direct measurement of $F_L$ (or $R$) at fixed ($x,Q^2$) requires data
at two different values of $y$ i.e. two different values of the CM
energy ($Q^2=sxy$). Since HERA operates at fixed CM energy, the value of
$F_L$ is extracted in four steps. One assumes that $F_2$ is well
described by NLO-QCD. Thus $F_2$ is determined from a NLO-QCD fit at a low
value of $y$ ($< 0.35$) where the $F_L$ contribution is negligible. This is 
then extrapolated to high values of $y$
($\simeq 0.7$) and then this value is used to subtract the $F_2$
contribution from the cross section at $y\simeq 0.7$. A typical
extraction of $F_L$ is shown in the figure 7 below.
\begin{figure}[htb]
\figurebox{}{8truecm}[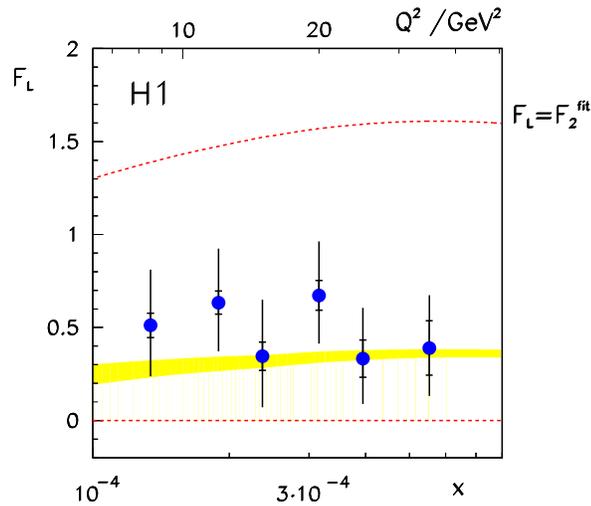]
\caption{A typical extraction of $F_L$ at $y=0.7$. The shaded band
corresponds to the allowed range in $F_L$ from parton distributions
determined from QCD as explained in the text. The dashed lines
correspond to limits of $F_L=0$  and $F_L=F_2$. }
\end{figure}

Before we discuss issues of more exclusive events in order to
differentiate between various evolution equations like DGLAP and BFKL,
some discussion on the details of the phenomenology of these equations
are in order. We have seen earlier the NLO-DGLAP indeed seems to work
very well in describing the data down to $2 GeV^2$. Do we therefore even
need any other evolution equations like BFKL to describe the data on
$F_2$. The answer is yes, because the 'perfect' fit of DGLAP to the data
depends crucially on the input parton densities at low $Q^2$. To
understand this fact, we need to compare the LO and NLO parton input
densities. Thus the fact that the GRV parametrisation works so well is
crucially dependent on the fact that the gluon densities have to be varied
a lot. This is clear from figure 8 where the input gluon densities for
LO and NLO are shown in the GRV parametrisation.
\begin{figure}[htb]
\figurebox{}{8truecm}[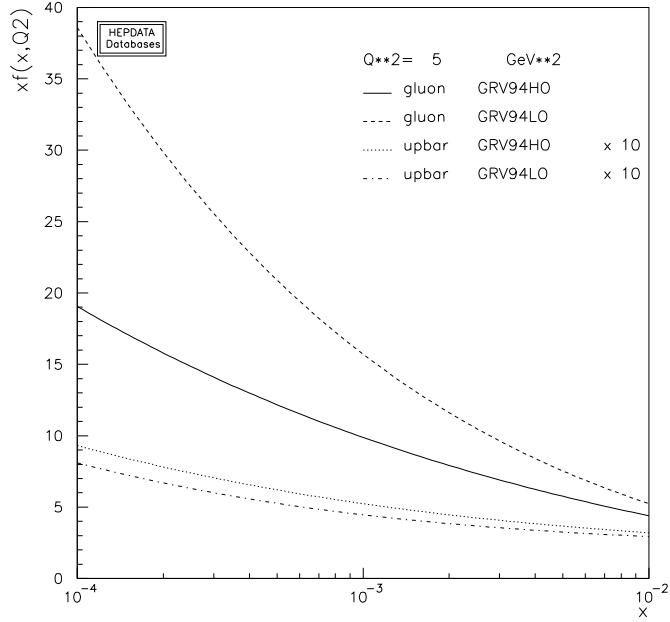]
\caption{Input gluon and $u$ quark densites at LO and NLO for GRV
parametrisation}
\end{figure}
The evolution of $F_2$ is given by 
\begin{equation}
\frac{dF_2}{d\ln Q^2}\sim P_{qg}\otimes g
\end{equation}
where
\begin{equation}
P_{qg}= \alpha_sP_{qg}^{(0)}\left[1+2.2 {3\alpha_s\over\pi}{1\over x}
+\ldots\right]
\end{equation}
Notice that the LO term is flat while the NLO term is steep in $x$. Thus
a stable evolution of $F_2$ requires that the NLO steepness of $P_{qg}$
has to be compensated by a gluon density which is {\em less steep} at
NLO than LO. This is clear from Figure 8 and tells us that there is a
strong correlation between $x$-shapes of $P_{qg}$ and $g$, implying that
fixed order DGLAP for $F_2$ needs extreme flexibility of parton density
parametrisations to offset the large NLO perturbative corrections at
small $x$. The NLO steepness of $P_{qg}$ is in fact the lowest order
manifestation of NLO BFKL in the quark channel.

Recently progress has been reported in calculating the
next-to-leading-log (NLL) contributions to the BFKL equation. Details
may be found in \cite{nll-bfkl}. Preliminary calculations seems to
indicate that NLL contributions stem the leading order rise of the
structure function ($x^{-\lambda}$ with $\lambda\simeq 0.5$) to 
$\lambda\simeq 2.65\alpha_s(Q)(1-c\alpha_s(Q))$ with $c=3.5$. 

The lesson to learn from all this is that we need detailed
investigations into the phenomenological and conceptual issues regarding
high energy log summations and more accurate signatures for BFKL
evolution in the hadronic final state. The selection criteria for these
have to be such that they suppress the phase space for DGLAP evolution
and maximise it for BFKL evolution. 
\section{Forward Jet Events}

One of these methods - the forward jet measurement has already been
discussed. Preliminary results from H1 are shown in Figure 9
and show a distinct preference for BFKL evolution, and is not reproduced
by the DGLAP Monte Carlos. 

\begin{figure}[htb]
\figurebox{}{8truecm}[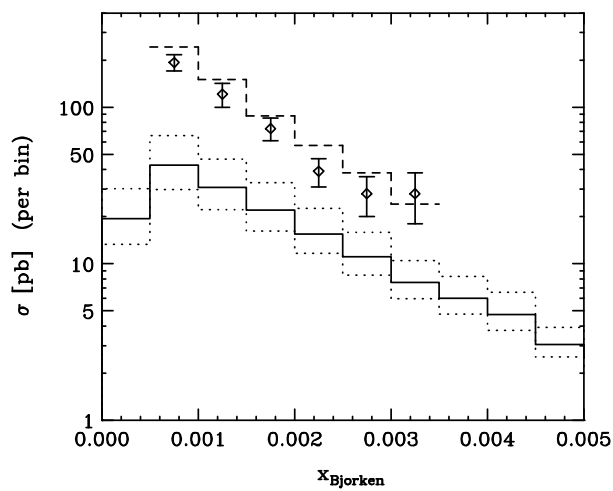]
\caption{Forward jet cross section at HERA as a function of $x$. The
solid histogram is the NLO MEPJET (DGLAP) result for the scale choice
$\mu_R^2=\mu_F^2=\xi(0.5\sum k_T)^2$ with $\xi=1$.
The two dotted histograms show the uncertainty of the NLO prediction,
corresponding
to a variation of $\xi$ between 0.1 and 10. The BFKL result of
Bartels et al ~\protect\cite{bartels} is shown as the dashed histogram. The 
data points are the new H1 measurements~\protect\cite{H1-forward}}
\end{figure}
The situation however is still far from clear. There are few theoretical
calculations in resummed DGLAP. Most predictions for the final state
observables are derived from Monte Carlo models whose drawback is their
complexity and flexibility to model hadronization. This makes it
difficult to pin down which feature of the theoretical input is being
tested when compared with the data for the various Monte Carlos in the
market like MEPJET, HERWIG, LEPTO etc. On the other hand no Monte Carlo
exist yet for BFKL evolution though some attempts have been made in that
direction by ARIADNE. For a detailed discussion of forward jet events
and its relevance to BFKL phenomenology, see \cite{delduca}.

\section{Single Particle Spectrum at large $p_T$}
Many of the ambiguities inherent in the forward jet event calculations
mentioned above can be overcome by considering single particle emission
at relatively large $p_T$ in the central region. Such an event is more
immune to hadronization effects and more directly reflects $\ln k_T^2$
diffusion from the BFKL ladder. The theoretical calculation has been
done in \cite{kwie} and preliminary data is shown in figure 10 for one
value of $x$. 
\begin{figure}[htb]
\figurebox{}{8cm}[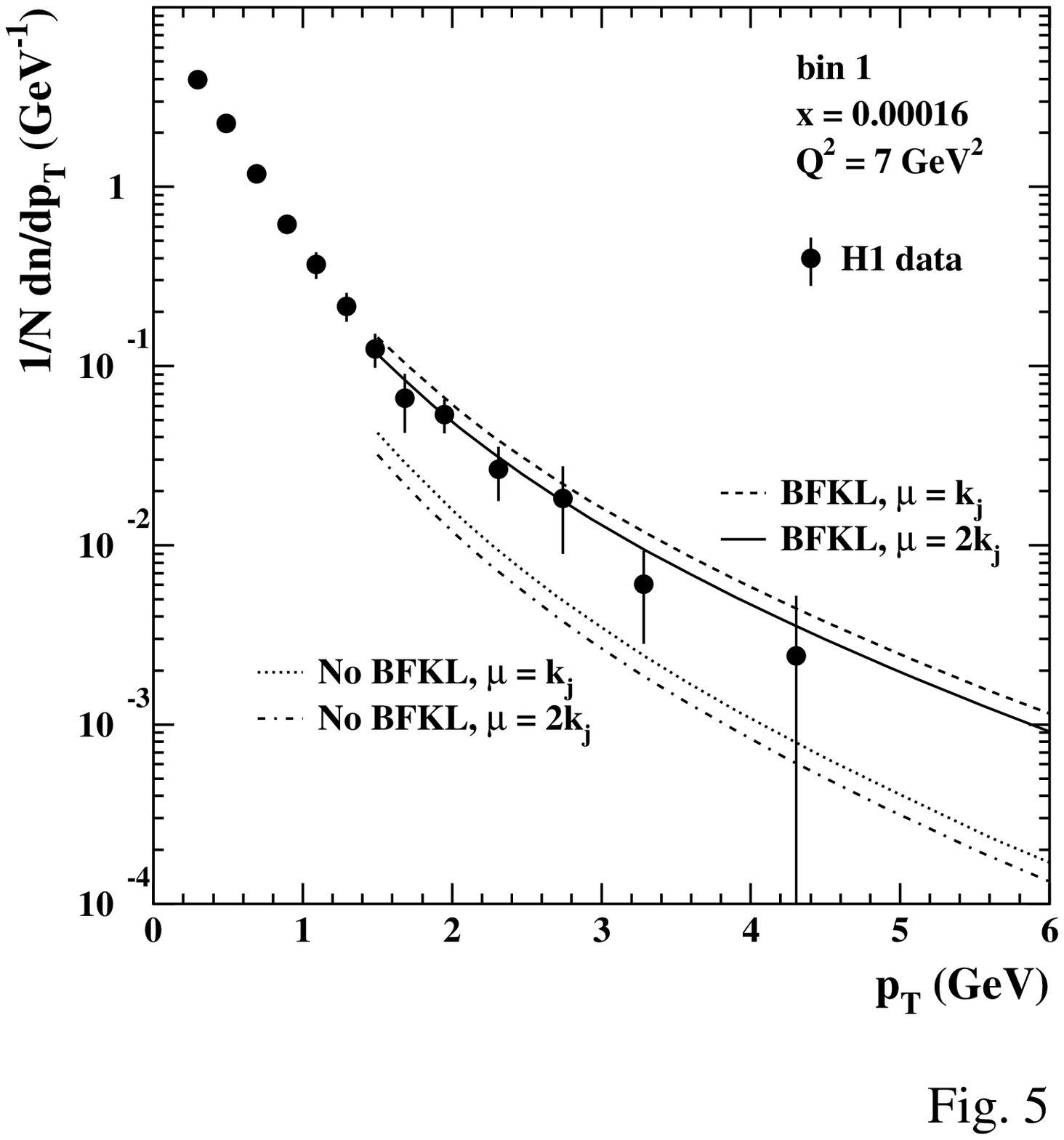]
\caption{The transverse momentum spectrum of charged
particles ($\pi^+, \pi^-, K^+, K^-$) in the pseudorapidity interval
$0.5 < \eta < 1.5$ in the virtual photon-proton centre-of-mass
frame.  The results are shown for kinematic bin 1 with the central values
$x = 1.6 \times 10^{-4}$ and $Q^2 = 7$ GeV$^2$. The continuous
and the dashed curve show the spectra obtained with $\Phi_i$ and $f$
calculated from the BFKL equation. They only differ in the choice
of fragmentation scale: for the continuous curve the fragmentation
functions were evaluated at scale $\mu^2 = (2k_j)^2$ and for the
dashed curve at scale $\mu^2 = k_j^2$. When BFKL radiation is neglected in the
calculation of the $p_T$ spectra, i.e.\ when the quark box approximation
$\Phi_i = \Phi_i^{(0)}$ is used and strong ordering at the gluon
vertex is assumed, then the
dash-dotted and dotted curves are obtained. The fragmentation
functions were evaluated at scales $2k_j$ and $k_j$ respectively. The data
points shown are from the H1 measurement of the charged particle spectra}
\end{figure}
There seems to be some evidence of $\ln (1/x)$ effects typical of BFKL
evolution and diffusion. However it would be useful to compare this with
the full fixed order calculation which would allow a clearer
distinction between the different predictions. 
The full DGLAP calculation for this process has not yet been
done and is presently under study.\cite{bg}.
\section{Conclusions}
In this talk we have concentrated on the various measurements of the
structure funcions $F_2$ as also $F_L$. We have compared the $F_2$
measurements with various theoretical calculations corresponding to
different evolution equations (DGLAP, BFKL ..). We have argued that the
$F_2$ measurements alone, being 'too inclusive' are not sufficient to
distinguish between different resummation programs. We present various
other proposals made in the literature for 'less inclusive' events - in
particular we discuss forward jet and single particle $p_T$ spectrum as
signatures for distinguishing fixed order calculations like DGLAP with
BFKL and indicate that while there seems to be evidence of the $\ln 1/x$
type of evolution and $k_T$ diffusion typical of BFKL, more theoretical
calculations and more precise experimental results are needed before any
conclusive evidence is obtained of the presence of BFKL effects.

\acknowledgement{I would like to thank the organisers of WHEPP-5 for
inviting me to the workshop and for providing an excellent atmosphere
for work and interaction.}
\begin{numberedbiblio}{99}
\bibitem{h1}
H1 Collaboration; T. Ahmed et al., Nucl. Phys. {\bf B 429} (1994) 477;
S. Aid et al., Nucl. Phys. {\bf B 470} (1996) 3.
\bibitem{zeus}
ZEUS Collaboration; M. Derrick et al., Phys. Lett. {\bf B315} (1993) 481;
Z. Phys {\bf C69}, (1996) 607
\bibitem{dola}
A. Donnachie and P. V. Landshoff, Nucl. Phys {\bf B 231} (1983) 189;
 ibid. {\bf B244} (1984) 322; ibid. {\bf B267} (1986) 690; Phys. Lett.
 {\bf B296} (1992) 227; ibid {\bf B202} (1988) 131.
\bibitem{grv}
M. Gluck, E. Reya and A. Vogt, Z. Phys. {\bf C 67} (1995) 433.
\bibitem{mrs}
A. D. Martin, W. J. Stirling and R. G. Roberts, Phys. Rev. {\bf D 50}
(1994) 6734.
\bibitem{dok}
Y. L. Dokshitzer, Sov. Phys. JETP {\bf 46} (1977) 641 
\bibitem{glap}	
V. N. Gribov and L. N. Lipatov, Sov. J. Nucl. Phys. {\bf 15} (1972) 438, 675; 
G.  Altarelli and G. Parisi, Nucl. Phys. {B 126} (1977) 298.
\bibitem{bfkl}
V. Fadin, E. Kuraev, L. Lipatov, Sov. Phys. JETP {\bf 44} (1976) 443;
ibid. {\bf 45} (1977) 199; Y. Balitski and L. Lipatov, Sov. J. Nucl.
Phys. {\bf 28} (1978) 822.
\bibitem{mueller}
A. H. Mueller, Nucl. Phys. {\bf B} (Proc. Suppl.), {\bf 18C},(1990) 125 ;
J. Phys. {G 17}, (1991) 1443
\bibitem{bf}
R. D. Ball and S. Forte, Phys. Lett. {\bf B 335} (1994) 77; Phys.
Lett. {\bf B 336} (1994) 77.
\bibitem{derujula}
A. De Rujula et al., Phys. Rev. {\bf D10} (1974) 1649.
\bibitem{foster}
Deep Inelastic Scattering at HERA - hep-ex/9712030
\bibitem{coopersarkar}
A. M. Cooper-Sarkar, R. C. E. Devenish and A. De Roeck - 
Structure Functions of the Nucleon and their Interpretation -
hep-ph/9712301
\bibitem{H1-forward}
H1 Collaboration, S. Aid et al., Phys. Lett. {\bf B356}, (1995) 118 .
\bibitem{nll-bfkl}
There are a large number of papers on the NLL corrections to the BFKL LL
result. An overview of these papers and the references can however be
found in V. Del Duca, BFKL: a minireview - hep-ph/9706552 
\bibitem{delduca}
V. Del Duca, BFKL in forward jet production - hep-ph/9707348.
\bibitem{bartels}
J.~Bartels et al., Phys. Lett. {\bf B384} (1996) 300 [hep-ph/9604272].
\bibitem{kwie}
J. Kwiecinski, S. C. Lang and A. D. Martin, 
Single particle spectra in deep inelastic scattering as a probe of small
$x$ dynamics - hep-ph/9707240
\bibitem{bg}
NLO-DGLAP calculation for single particle spectrum - R. Basu and R. M.
Godbole - in preparation
\end{numberedbiblio}
\end{document}